\documentclass[twocolumn,showpacs,pra]{revtex4-1}%
\usepackage{amsfonts}
\usepackage{amsmath}
\usepackage{amssymb}
\usepackage{subfigure}
\usepackage{graphicx}
\usepackage{txfonts}%
\providecommand{\U}[1]{\protect\rule{.1in}{.1in}}

\begin{document}
\title{Focusing RKKY interaction by graphene P-N junction}
\author{Shu-Hui Zhang$^{1}$}
\author{Jia-Ji Zhu$^{3}$}
\author{Wen Yang$^{1}$}
\email{wenyang@csrc.ac.cn}
\author{Kai Chang$^{2,4}$}
\email{kchang@semi.ac.cn}
\affiliation{$^{1}$Beijing Computational Science Research Center, Beijing 100193, China}
\affiliation{$^{2}$SKLSM, Institute of Semiconductors, Chinese Academy of Sciences, P.O.
Box 912, Beijing 100083, China}
\affiliation{$^{3}$Institute for quantum information and spintronics, School of Science,
Chongqing University of Posts and Telecommunications, Chongqing 400065, China}
\affiliation{$^{4}$Synergetic Innovation Center of Quantum Information and Quantum Physics,
University of Science and Technology of China, Hefei, Anhui 230026, China}

\begin{abstract}
The carrier-mediated RKKY interaction between local spins plays an important
role for the application of magnetically doped graphene in spintronics and
quantum computation. Previous studies largely concentrate on the influence of
electronic states of uniform systems on the RKKY\ interaction. Here we reveal
a very different way to manipulate the RKKY interaction by showing that the
anomalous focusing -- a well-known electron optics phenomenon in graphene P-N
junctions -- can be utilized to refocus the massless Dirac electrons emanating
from one local spin to the other local spin. This gives rise to rich spatial
interference patterns and symmetry-protected non-oscillatory RKKY\ interaction
with a strongly enhanced magnitude. It may provide a new way to engineer the
long-range spin-spin interaction in graphene.

\end{abstract}

\pacs{73.40.Lq, 75.30.Hx, 72.80.Vp, 73.23.Ad}
\maketitle

\section{Introduction}

Graphene \cite{CastroRMP2009}, a single atomic layer of graphite, is featured
by ultra-high mobility and electrical tunability of carrier density and hence
provides an attractive platform for studying the unique electron optics of
Dirac fermions owing to its gapless and linear dispersion. Cheianov \textit{et
al.} \cite{CheianovScience2007} proposed the interesting idea that an
interface between electron (N)-doped and hole (P)-doped regions in graphene
can focus an electron beam, which may lead to the realization of an electronic
analog of the Veselago lens in optics
\cite{VeselagoSPU1968,PendryPRL2000,ZhangNatMater2008,PendryScience2012}. This
anomalous focusing effect has motivated many new ideas and device concepts
\cite{CsertiPRL2007,GarciaPomarPRL2008,MoghaddamPRL2010,SilveirinhaPRL2013,ZhaoPRL2013}%
. Very recently, this effect was observed experimentally
\cite{LeeNatPhys2015,ChenScience2016}, which paves the way for realizing
electron optics based on graphene P-N junctions (PNJs). A common feature of
these works is that they concentrate on focusing the electrons themselves,
leaving its potential applications to other fields unexplored.

In this work, we explore a very different direction by showing that the
anomalous focusing effect could be utilized to manipulate the carrier-mediated
Rudermann-Kittel-Kasuya-Yosida (RKKY) interaction
\cite{RudermanPR1954,KasuyaPTP1956,YosidaPR1957} between magnetic moments
(spins), with potential applications in spintronics
\cite{WolfScience2001,ZuticRMP2004,MacDonaldNatMater2005}, scalable quantum
computation \cite{LossPRA1998,TrifunovicPRX2012}, and majorana fermion physics
\cite{KlinovajaPRL2013} as the RKKY\ interaction enables long-range
correlation between spatially separated local spins
\cite{PiermarocchiPRL2002,CraigScience2004,RikitakePRB2005,FriesenPRL2007,SrinivasaPRL2015}%
, a crucial ingredient in these developments. In recent years, a lot of
efforts have been devoted to characterizing the RKKY\ interaction in different
2D \textit{uniform} systems such as two-dimensional electron gases
\cite{FischerPRB1975,Beal-MonodPRB1987}, graphene
\cite{BreyPRL2007,SaremiPRB2007,BlackSchafferPRB2010,SherafatiPRB2011,SherafatiPRB2011a,UchoaPRL2011,KoganPRB2011}%
, and the surface of topological insulators
\cite{LiuPRL2009,ZhuPRL2011,AbaninPRL2011}. There are also many interesting
schemes to manipulate the RKKY\ interaction
\cite{InosovPRL2009,ZhuPRB2010,NeelPRL2011,YaoPRL2014,HolubAPL2004,XiuNatMater2010,NieNatComm2016,BouhassouneNatComm2014,PowerPRB2012}%
\textbf{. } Recent experimental advances further enable the RKKY interaction
to be mapped out with atomic-scale resolution from single-atom magnetometry
using scanning tunnelling spectroscopy
\cite{MeierScience2008,ZhouNatPhys2010,KhajetooriansNatPhys2012}.\textbf{
}However, in a $d$-dimensional uniform system, the RKKY\ interaction usually
decays at least as fast as $1/R^{d}$. This rapid decay -- a common feature of
the previous studies mentioned above -- makes the RKKY\ interaction very
short-ranged and may hinder its applications. This motivates growing interest
in modifying the long-range behavior of the RKKY interaction, e.g., the
$1/R^{3}$ long-range decay in \textit{undoped} graphene can be changed by
thermal excitation \cite{KlierPRB2015} and even be slowed down by
electron-electron interactions \cite{Black-SchafferPRB2010}. Here we show that
the graphene PNJ allows the diverging electron beams emanating from one local
spin to be refocused onto the other local spin, thus the electron-mediated
RKKY interaction between these two local spins can be strongly enhanced and
tuned beyond the $1/R^{d}$ limit of non-interacting uniform systems. The
graphene PNJ also gives rise to symmetry-protected non-oscillatory RKKY
interaction as a function of the distance, in sharp contrast to the
\textquotedblleft universal\textquotedblright\ oscillation of the
RKKY\ interaction in uniform systems with a finite carrier concentration. This
may provide a new way for engineering the correlation between spatially
separated local spins for their applications in spintronics and quantum computation.

This paper is organized as follows. In Sec. II, we explain intuitively how to
utilize the graphene PNJ to manipulate the RKKY\ interaction and highlight the
symmetry-protected non-oscillatory RKKY\ interaction. Then in Sec. III, we
perform numerical simulation based on the tight-binding model to demonstrate
this all-electrical manipulation and discuss the experimental feasibility.
Finally, we present a brief summary in Sec. IV.

\section{RKKY$\ $interaction in graphene P-N junction: physical picture}

Let us consider two local spins $\hat{\mathbf{S}}_{1}$ (located at
$\mathbf{R}_{1}$) and $\hat{\mathbf{S}}_{2}$ (located at $\mathbf{R}_{2}$)
coupled to the spin density $\hat{\mathbf{s}}(\mathbf{x})\equiv\hat
{\mathbf{s}}\delta(\hat{\mathbf{r}}-\mathbf{x})$ of itinerant carriers via the
exchange interaction $\hat{V}_{\mathrm{ex}}=-J_{0}\hat{\mathbf{S}}_{1}%
\cdot\hat{\mathbf{s}}(\mathbf{R}_{1})-J_{0}\hat{\mathbf{S}}_{2}\cdot
\hat{\mathbf{s}}(\mathbf{R}_{2}),$ where $\hat{\mathbf{s}}$ ($\hat{\mathbf{r}%
}$) is the carrier spin (position) operator. The total Hamiltonian of the
coupled system is the sum of $\hat{V}_{\mathrm{ex}}$ and the carrier
Hamiltonian $\hat{H}$. The carrier-mediated RKKY interaction originates from
the local excitation of carrier spin density fluctuation by one local spin and
its subsequent propagation to the other spin. At zero temperature, the
effective RKKY\ interaction between the local spins is obtained by eliminating
the carrier degree of freedom through second-order perturbation theory as
\cite{RudermanPR1954,KasuyaPTP1956,YosidaPR1957,BrunoPRB1995} $\hat
{H}_{\mathrm{RKKY}}=\sum_{\alpha\beta=x,y,z}J_{\alpha\beta}\hat{S}_{1}%
^{\alpha}\hat{S}_{2}^{\beta}$, where the RKKY\ range function
\begin{equation}
J_{\alpha\beta}=-\frac{J_{0}^{2}}{\pi}\int_{-\infty}^{E_{F}}\operatorname{Im}%
\mathrm{Tr}[\hat{s}_{\alpha}\hat{G}(\mathbf{R}_{1},\mathbf{R}_{2};E)\hat
{s}_{\beta}\hat{G}(\mathbf{R}_{2},\mathbf{R}_{1};E)]dE, \label{J_RKKY}%
\end{equation}
$E_{F}$ is the Fermi energy of the carriers, $\mathrm{Tr}$ traces over the
carrier spin, and $\hat{G}(\mathbf{r},{\mathbf{r}}_{0};E)\equiv\langle
\mathbf{r}|(E+i0^{+}-\hat{H})|\mathbf{r}_{0}\rangle$ is the unperturbed (i.e.,
in the absence of the local spins) propagator of the carriers in real space.
In general, $\hat{G}(\mathbf{r},{\mathbf{r}}_{0};E)$ is still an operator
acting on the carrier spin degree of freedom. In a $d$-dimensional uniform
system, the carriers excited by the first local spin at $\mathbf{R}_{1}$
propagate towards the second local spin at $\mathbf{R}_{2}$ in the form of an
outgoing wave $\hat{G}({\mathbf{R}_{2}},{\mathbf{R}_{1}},E)\sim e^{ikR}%
/R^{(d-1)/2}$, where $R\equiv|\mathbf{R}_{2}-\mathbf{R}_{1}|$, $k$ is a
characteristic wave vector of the carriers with energy $E$, and the
denominator $R^{(d-1)/2}$ ensures the conservation of probability current. The
integration over the energy in Eq. (\ref{J_RKKY}) yields another factor $1/R$
from the oscillating phase factor $e^{ikR}$, so $J_{\alpha\beta}\propto
1/R^{d}$. This provides a rough explanation for the \textquotedblleft
universal\textquotedblright\ $1/R^{d}$ decay of the RKKY interaction, as
discovered in a great diversity of materials by previous studies. It also
reveals a very different way -- tailoring the carrier propagation and
interference -- to manipulate the RKKY\ interaction beyond this constraint, as
opposed to previous studies that exploit the electronic states and energy band
structures of different uniform materials. The anomalous focusing effect in
graphene PNJs \cite{CheianovScience2007} provides a paradigmatic example for
this manipulation.

\begin{figure}[t]
\includegraphics[width=\columnwidth,clip]{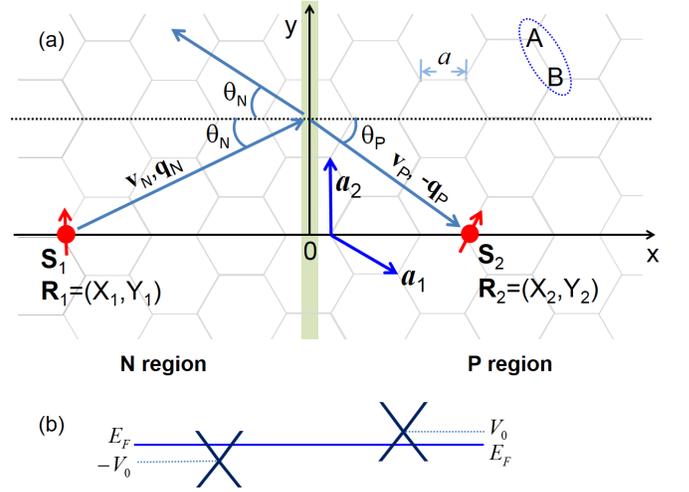}\caption{(a) Graphene
P-N junction at $x=0$, with two localized spins $\mathbf{S}_{1}$ (in the N
region) and $\mathbf{S}_{2}$ (in the P region). The unit cell of graphene
(dashed ellipse) consists of one atom on sublattice $A$ and one atom on
sublattice $B$. (b) Dirac cones of the N region and P region relative to the
Fermi energy. $\mathbf{v}_{\mathrm{N}}\parallel\mathbf{q}_{N}$ and
$\mathbf{v}_{\mathrm{P}}\parallel(-\mathbf{q}_{P})$ are the group velocities
of the incident and transmission waves, respectively.}%
\label{G_PNJ}%
\end{figure}

As shown in Fig. \ref{G_PNJ}(a), the honeycomb lattice of graphene consists of
two sublattices (denoted by $A$ and $B$) and each unit cell contains two
carbon atoms (or $\pi_{z}$-orbitals), one on each sublattice. Let us use
$\mathbf{R}$ to denote the location of each carbon \textit{atom},
$|\mathbf{R}\rangle$ for the corresponding orbital, and $s_{\mathbf{R}}$ ($=A$
or $B$) for the sublattice on which $\mathbf{R}$ locates. For carriers in the
graphene PNJ, the tight-binding Hamiltonian is the sum of $\hat{H}_{0}$ for
uniform graphene and $\hat{V}_{\mathrm{J}}$ for the on-site junction
potential:
\begin{subequations}
\label{HAMIL_TB}%
\begin{align}
\hat{H}  &  =\hat{H}_{0}+\hat{V}_{\mathrm{J}},\label{H0}\\
\hat{H}_{0}  &  =-t\sum_{\langle\mathbf{R},\mathbf{R}^{\prime}\rangle
}|\mathbf{R}\rangle\langle\mathbf{R}^{\prime}|+h.c.,\\
\hat{V}_{\mathrm{J}}  &  =\sum_{\mathbf{R}}V_{\mathbf{R}}|\mathbf{R}%
\rangle\langle\mathbf{R}|,
\end{align}
where $\langle\mathbf{R},\mathbf{R}^{\prime}\rangle$ denotes nearest
neighbors, $t\approx3$ eV is the nearest-neighbor hopping \cite{CastroRMP2009}%
, and $V_{\mathbf{R}}$ is equal to $-V_{0}$ ($+V_{0}$)\textbf{ }when
$\mathbf{R}$ locates in the left (right) of the junction [shaded stripe in
Fig. \ref{G_PNJ}(a)] with $V_{0}\geq0$. As shown in Fig. \ref{G_PNJ}(b), the
zero point of energy is chosen such that the Dirac point in the left (right)
of the junction lies at $-V_{0}$ ($+V_{0}$). Uniform graphene corresponds to
$V_{0}=0$, while nonzero $V_{0}$ corresponds to a junction, e.g., N-N (P-P)
junction corresponds to $E_{F}>V_{0}$ ($E_{F}<V_{0}$). Here we consider the
P-N junction (PNJ), corresponding to $E_{F}\in\lbrack-V_{0},+V_{0}]$. In
uniform graphene $(V_{0}=0$), the doping is determined by $E_{F}$. In graphene
PNJ, the electron doping in the N region (left) is $V_{0}+E_{F}$, while the
hole doping in the P region (right) is $V_{0}-E_{F}$, e.g., $E_{F}=0$
corresponds to the electron doping in the N region being equal to the hole
doping in the P region.

Due to the absence of spin-orbit coupling in the carrier Hamiltonian $\hat{H}%
$, the carrier-mediated RKKY interaction at zero temperature between one local
spin $\hat{\mathbf{S}}_{1}$ at $\mathbf{R}_{1}$\ (sublattice $s_{\mathbf{R}%
_{1}}$) in the N\ region and another local spin $\hat{\mathbf{S}}_{2}$ at
$\mathbf{R}_{2}$ (sublattice $s_{\mathbf{R}_{2}}$) in the P region [see Fig.
\ref{G_PNJ}(a)] assumes the isotropic Heisenberg form
\cite{SherafatiPRB2011,SherafatiPRB2011a}:\textbf{ }$\hat{H}_{\mathrm{RKKY}%
}=J\hat{\mathbf{S}}_{1}\cdot\hat{\mathbf{S}}_{2}$, where the range function%
\end{subequations}
\begin{equation}
J=-\frac{J_{0}^{2}}{2\pi}\int_{-\infty}^{E_{F}}\operatorname{Im}%
G^{2}(\mathbf{R}_{2},\mathbf{R}_{1},E)dE \label{JS1S2}%
\end{equation}
is determined by the unperturbed propagator (i.e., in the absence of the local
spins) of the carriers from $\mathbf{R}_{1}$ to $\mathbf{R}_{2}$:%
\[
G(\mathbf{R}_{2},\mathbf{R}_{1},E)\equiv\langle\mathbf{R}_{2}|(E+i0^{+}%
-\hat{H})|\mathbf{R}_{1}\rangle.
\]
In arriving at Eq. (\ref{JS1S2}), we have used $G(\mathbf{R}_{2}%
,\mathbf{R}_{1},E)=G(\mathbf{R}_{1},\mathbf{R}_{2},E)$ due to the
time-reversal invariance of the graphene Hamiltonian $\hat{H}$. Note that the
propagator $G(\mathbf{R}_{2},\mathbf{R}_{1},E)$ and hence the RKKY range
function $J$ are very sensitive to the sublattices on which $\mathbf{R}_{1}$
and $\mathbf{R}_{2}$ locate (i.e., $s_{\mathbf{R}_{1}}$ and $s_{\mathbf{R}%
_{2}}$), e.g., when $\mathbf{R}_{2}$ moves from an atom on the $A$ sublattice
($s_{\mathbf{R}_{2}}=A$) to a \textit{neighboring} atom on the $B$ sublattice
($s_{\mathbf{R}_{2}}=B$) [see Fig. \ref{G_PNJ}(a)], the propagator and hence
the RKKY\ interaction may change significantly.

Now we discuss how the anomalous focusing effect in graphene PNJs
\cite{CheianovScience2007} can be utilized to manipulate the carrier
propagator and hence the RKKY\ interaction beyond the \textquotedblleft
universal\textquotedblright\ $1/R^{d}$ long-range decay as encountered in
previous studies. Ever since the pioneering work of Cheianov \textit{et al.}
\cite{CheianovScience2007}, there have been many studies on the anomalous
focusing effect, either based on the classical analogy to light propagation in
geometric optics or based on the scattering of the electron wave functions.
Below, we provide a physically intuitive analysis on how the graphene PNJ
focuses the carrier propagator and hence the RKKY\ interaction based on the
continuum model of graphene\textbf{ }\cite{CastroRMP2009}. The purpose is to
provide a qualitative picture for the focusing of the RKKY\ interaction and
establish its effectiveness for an arbitrary direction of the P-N interface.

\subsection{Anomalous focusing of carrier propagator: continuum model}

The low-energy physics of graphene is described by two Dirac cones located at
$\mathbf{K}$ and $\mathbf{K}^{\prime}=-\mathbf{K}$, which form a Kramer pair.
For clarity, we first analyze the focusing effect based on the $\mathbf{K}%
$-valley continuum model, leaving the discussion including both valleys to the
end of this subsection. Using the band-edge Bloch functions $|\Phi
_{\mathbf{K},A}\rangle$ (from $\pi_{z}$-orbitals on the $A$ sublattice) and
$|\Phi_{\mathbf{K},B}\rangle$ (from $\pi_{z}$-orbitals on the $B$ sublattice)
of the $\mathbf{K}$ valley as the basis, the continuum model for the
$\mathbf{K}$ valley reads \cite{CastroRMP2009},
\begin{equation}
\hat{h}=v_{F}\boldsymbol{\hat{\sigma}}\cdot\mathbf{\hat{p}}+\mathrm{sgn}%
(x)V_{0}, \label{HAMIL_CONTINUUM}%
\end{equation}
where $\mathbf{\hat{p}}$ is the momentum relative to the $\mathbf{K}$ valley and $v_F$ is the Fermi velocity.
Note that the continuum model regards the two atoms of the same unit cell to
locate at the same spatial point, so each spatial point contains two
sublattices/orbitals and the Hamiltonian $\hat{h}$ is a $2\times2$ matrix.
Correspondingly, the carrier propagator from $\mathbf{R}_{1}$ to
$\mathbf{R}_{2}$ is also a 2$\times$2 matrix: $\mathbf{g}(\mathbf{R}%
_{2},\mathbf{R}_{1},E)\equiv\langle\mathbf{R}_{2}|(E+i0^{+}-\hat
{h})|\mathbf{R}_{1}\rangle$, e.g., its $(B,A)$ matrix element gives the
carrier propagator from the sublattice $A$ at $\mathbf{R}_{1}$ to the
sublattice $B$ at $\mathbf{R}_{2}$.

For uniform graphene, the $\mathbf{K}$-valley continuum model [Eq.
(\ref{HAMIL_CONTINUUM}) with $V_{0}=0$] leads to a massless Dirac spectrum
$E_{\pm}(\mathbf{q})\equiv\pm v_{F}|\mathbf{q|}$ and chiral eigenstates
$|\pm,\mathbf{q}\rangle=e^{i\mathbf{q}\cdot\mathbf{r}}|u_{\pm}(\mathbf{q}%
)\rangle$, where $|u_{\pm}(\mathbf{q})\rangle$ is the two-component spinor for
the sublattice degrees of freedom. The conduction (valence) band state
$|+,\mathbf{q}\rangle$ ($|-,\mathbf{q}\rangle$) has a group velocity
$\mathbf{v}(\mathbf{q})=v_{F}\mathbf{q}/|\mathbf{q}|$ [$-\mathbf{v}%
(\mathbf{q})\equiv-v_{F}\mathbf{q}/|\mathbf{q}|$] parallel (anti-parallel) to
the momentum $\mathbf{q}$. The RKKY\ interaction is usually dominated by the
contributions from carriers near the Fermi surface [the energy integral in Eq.
(\ref{JS1S2}) merely produces a multiplicative factor $\propto1/R$], so we
focus on the carrier propagator on the Fermi level $E_{F}$. For $E_{F}>0$, the
Fermi momentum is $q_{F}\equiv E_{F}/v_{F}$, and the right-going eigenstates
$|+,\mathbf{q}\rangle$ on the Fermi contour are characterized by the momentum
$\mathbf{q}\equiv((q_{F}^{2}-q_{y}^{2})^{1/2},q_{y})$. The 2$\times$2
propagator from $\mathbf{R}_{1}$ to $\mathbf{R}_{2}$ (on the right of
$\mathbf{R}_{1}$) in \textit{uniform} graphene can be expressed in terms of
these eigenstates as
\begin{equation}
\mathbf{g}_{\mathrm{uniform}}(\mathbf{R}_{2},\mathbf{R}_{1},E_{F}%
)=\int_{-\infty}^{\infty}\frac{dq_{y}}{2\pi}\frac{|u_{+}(\mathbf{q}%
)\rangle\langle u_{+}(\mathbf{q})|}{iv_{x}(\mathbf{q})}e^{i\mathbf{q}%
\cdot(\mathbf{R}_{2}-\mathbf{R}_{1})}. \label{GF0}%
\end{equation}
The RKKY\ interaction in uniform graphene is given by Eq. (\ref{JS1S2}) with
$G(\mathbf{R}_{2},\mathbf{R}_{1},E)$ replaced by the $(s_{\mathbf{R}_{2}%
},s_{\mathbf{R}_{1}})$ matrix element of $\mathbf{g}_{\mathrm{uniform}%
}(\mathbf{R}_{2},\mathbf{R}_{1},E_{F})$.

For the graphene PNJ described by the $\mathbf{K}$-valley continuum model in
Eq. (\ref{HAMIL_CONTINUUM}), the first local spin $\mathbf{S}_{1}$ in the N
region excites a series of outgoing plane wave eigenstates on the Fermi
contour, but only the right-going eigenstates, i.e., $|+,\mathbf{q}_{N}%
\rangle$ with momentum $\mathbf{q}_{N}\equiv((q_{N}^{2}-q_{y}^{2})^{1/2}%
,q_{y})$, can transmit across the P-N\ interface, becomes a right-going
eigenstate $|-,\mathbf{q}_{P}\rangle$ with momentum $\mathbf{q}_{P}%
\equiv(-(q_{P}^{2}-q_{y}^{2})^{1/2},q_{y})$ on the Fermi contour of the P
region, and finally arrive at $\mathbf{S}_{2}$, where $q_{N}\equiv(V_{0}%
+E_{F})/v_{F}$ and $q_{P}\equiv(V_{0}-E_{F})/v_{F}$ are Fermi momenta in the N
and P regions, respectively. In terms of these local, right-going eigenstates
on the Fermi contours, the carrier propagator from $\mathbf{R}_{1}$ to
$\mathbf{R}_{2}$ is (see Appendix A):%
\begin{equation}
\mathbf{g}(\mathbf{R}_{2},\mathbf{R}_{1},E)=\int_{-\infty}^{\infty}%
\frac{dq_{y}}{2\pi}w(q_{y})\frac{|u_{-}(\mathbf{q}_{P})\rangle\langle
u_{+}(\mathbf{q}_{N})|}{iv_{x}(\mathbf{q}_{N})}e^{i\mathbf{q}_{P}%
\cdot\mathbf{R}_{2}}e^{-i\mathbf{q}_{N}\cdot\mathbf{R}_{1}}, \label{GF}%
\end{equation}
where $w(q_{y})$ is the transmission coefficient of the incident state
$|+,\mathbf{q}_{N}\rangle$ across the PNJ. The RKKY\ interaction in the
graphene PNJ is given by Eq. (\ref{JS1S2}) with $G(\mathbf{R}_{2}%
,\mathbf{R}_{1},E)$ replaced by the $(s_{\mathbf{R}_{2}},s_{\mathbf{R}_{1}})$
matrix element of $\mathbf{g}(\mathbf{R}_{2},\mathbf{R}_{1},E_{F})$.

\begin{figure}[t]
\includegraphics[width=\columnwidth,clip]{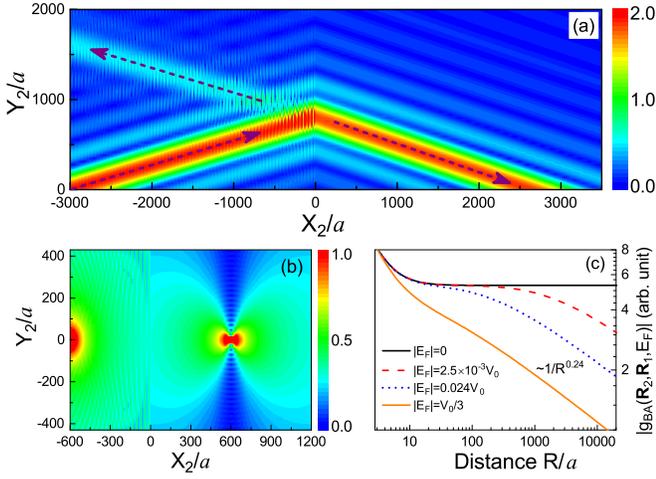}\caption{Anomalous
focusing {across a graphene PNJ at }$x=0${. }The propagator is the sum of the
contributions from all the wave packets characterized by different center
momenta on the Fermi contour. (a) Contribution of a \textit{single} wave
packet on the Fermi contour (whose center momentum has an incident angle
$\theta_{N}=15\operatorname{{{}^{\circ}}}$) to the propagator $|g_{BA}%
(\mathbf{R}_{2},\mathbf{R}_{1},E_{F}=0.03t)|$ vs. $\mathbf{R}_{2}=(X_{2}%
,Y_{2})$ on the $B$ sublattice, where $\mathbf{R}_{1}=(X_{1},Y_{1})$ is fixed
at $X_{1}=-3000a$ and $Y_{1}=0$ on the $A$ sublattice. The dashed arrows mark
the classical trajectory. (b) Propagator $|g_{BA}(\mathbf{R}_{2}%
,\mathbf{R}_{1},E_{F}=0)|$ vs. $\mathbf{R}_{2}$ on $B$ sublattice for fixed
$\mathbf{R}_{1}=(-601a,0)$ on $A$ sublattice. (c) Decay of the propagator
$|g_{BA}(\mathbf{R}_{2},\mathbf{R}_{1},E_{F})|$ along the $x$ axis (i.e.,
$Y_{1}=Y_{2}=0$) with increasing distance $R$, where $\mathbf{R}_{2}$ is on
the cusp of the caustics. The PNJ potential $V_{0}=t/2$ for (a) and $t/5$ for
(b)\ and (c). }%
\label{G_PROPAGATOR}%
\end{figure}

The key difference between the carrier propagator in the graphene PNJ [Eq.
(\ref{GF})] and that in uniform graphene [Eq. (\ref{GF0})] is the change of
the propagation phase factor from $e^{i\mathbf{q}\cdot(\mathbf{R}%
_{2}-\mathbf{R}_{1})}\equiv e^{i\phi_{0}(q_{y})}$ for uniform graphene to
$e^{i(\mathbf{q}_{P}\cdot\mathbf{R}_{2}-\mathbf{q}_{N}\cdot\mathbf{R}_{1}%
)}\equiv e^{i\phi_{\mathrm{NP}}(q_{y})}$ for the graphene PNJ. This change is
responsible for the anomalous focusing of the diverging carrier wave into a
converging one. For an intuitive analysis of this behavior, we discretize the
$q_{y}$ axis into grids $m\Delta$ ($m\in\mathbb{Z}$), where the spacing
$\Delta\ll$ size of the graphene Brillouin zone. Then Eq. (\ref{GF}) gives
$\mathbf{g}(\mathbf{R}_{2},\mathbf{R}_{1},E)=\sum_{m}\mathbf{g}^{(m)}%
(\mathbf{R}_{2},\mathbf{R}_{1},E)$ and $\mathbf{g}^{(m)}$ is the contribution
from the $q_{y}$ integral over the $m$th segment $[(m-1/2)\Delta
,(m+1/2)\Delta]$, corresponding to a wave packet characterized by the center
momentum $q_{y}=m\Delta$. In other words, the entire propagator is the sum of
contributions from all these wave packets characterized by different center
momenta $q_{y}$'s on the Fermi contour. The same analysis is applicable to the
propagator $\mathbf{g}_{\mathrm{uniform}}$ in uniform graphene [Eq.
(\ref{GF0})]. Since the integrand is the product of a slowly-varying part and
a rapidly oscillating propagation phase factor, the contribution from a given
wave packet characterized by the center momentum $q_{y}$ is appreciable only
when the propagation phase is stationary:$\ \partial_{q_{y}}\phi_{0}(q_{y})=0$
(for uniform graphene) or $\partial_{q_{y}}\phi_{\mathrm{NP}}(q_{y})=0$ (for
graphene PNJ). This first-order stationary phase condition determines the most
probable (or classical) trajectory of a wave packet emanating from
$\mathbf{R}_{1}$.

In uniform graphene, the classical trajectory of a \textit{given} wave packet
characterized by the center momentum $\mathbf{q}=((q_{F}^{2}-q_{y}^{2}%
)^{1/2},q_{y})$ on the Fermi contour [Eq. (\ref{GF0})] is a beam emanating
from $\mathbf{R}_{1}$ and going along the wave vector $\mathbf{q}$. The
classical trajectories of different wave packets on the Fermi contour form
many outgoing beams emanating from $\mathbf{R}_{1}$, which manifests the
diverging propagation of the carriers in uniform graphene and leads to
$\mathbf{g}_{\mathrm{uniform}}\propto1/R^{1/2}$. By contrast, in the graphene
PNJ, the classical trajectory of a \textit{given} wave packet characterized by
the center momentum $\mathbf{q}_{N}=((q_{N}^{2}-q_{y}^{2})^{1/2},q_{y})$ with
incident angle $\theta_{N}=\tan^{-1}(q_{y}/q_{N,x})$ consists of the incident
beam along $\mathbf{q}_{N}$, the reflection beam with a reflection angle
$\theta_{N}$, and the refraction beam with a refraction angle $\theta
_{P}\equiv\tan^{-1}(q_{y}/q_{P,x})$, as sketched in Fig. \ref{G_PNJ}(a) and
further visualized in Fig. \ref{G_PROPAGATOR}(a). Here the refraction angle
$\theta_{P}$ is determined by the Snell law $\sin\theta_{N}=n\sin\theta_{P}$
with a negative effective refractive index $n\equiv-(V_{0}-E_{F})/(V_{0}%
+E_{F})$ \cite{CheianovScience2007}.

Let us use $(X_{1},Y_{1})$ and $(X_{2},Y_{2})$ to denote the Cartesian
coordinates of $\mathbf{R}_{1}$ and $\mathbf{R}_{2}$, respectively. In the
graphene PNJ, when $\mathbf{R}_{2}$ locates on the caustics
\cite{CheianovScience2007}
\begin{equation}
(Y_{2}-Y_{1})^{2}=\frac{[X_{2}^{2/3}-(nX_{1})^{2/3}]^{3}}{n^{2}-1},
\label{CAUSTICS}%
\end{equation}
the wave packet going from $\mathbf{R}_{1}$ to $\mathbf{R}_{2}$ obeys not only
$\partial_{q_{y}}\phi_{0}(q_{y})=0$, but also $\partial_{q_{y}}^{2}\phi
_{0}(q_{y})=0$, so that its contribution to the propagator $\mathbf{g}%
(\mathbf{R}_{2},\mathbf{R}_{1},E_{F})$ is enhanced. The most interesting case
occurs at $E_{F}=0$ or equivalently $n=-1$. In this case, we have
$q_{N,x}=-q_{P,x}$, thus for $\mathbf{R}_{2}$ at the mirror image of
$\mathbf{R}_{1}$ about the PNJ, i.e., $\mathbf{R}_{2}=\mathbf{R}%
_{1}^{\mathrm{m}}\equiv(|X_{1}|,Y_{1})$, the phase $\phi_{\mathrm{NP}}(q_{y})$
vanishes for all $q_{y}$, so that the integrand in Eq. (\ref{GF}) no longer
suffers from the rapidly oscillating phase factor $e^{i\phi_{\mathrm{NP}%
}(q_{y})}$. This corresponds to constructive interference of all the
transmission waves at $\mathbf{R}_{1}^{\mathrm{m}}$ or equivalently perfect
focusing of the diverging electron beams emanating from $\mathbf{R}_{1}$ onto
$\mathbf{R}_{1}^{\mathrm{m}}$ \cite{CheianovScience2007}. This not only lead
to strong local enhancement of the propagator $\mathbf{g}(\mathbf{R}%
_{2},\mathbf{R}_{1},E_{F}=0)$ when $\mathbf{R}_{2}$ locates in the vicinity of
$\mathbf{R}_{1}^{\mathrm{m}}$ [see Fig. \ref{G_PROPAGATOR}(b)], but also makes
$\mathbf{g}(\mathbf{R}_{1}^{\mathrm{m}},\mathbf{R}_{1},E_{F}=0)$ independent
of the distance $R$ [black, solid line in Fig. \ref{G_PROPAGATOR}(c)], in
sharp contrast to the $1/R^{1/2}$ decay in uniform graphene. This distance
independent propagator can be attributed to the existence of a hidden symmetry
on the Fermi contours of the host material \cite{ZhangPRB2016}.\

When $E_{F}\neq0$, the anomalous focusing locally enhances the propagator from
$\mathbf{R}_{1}$ to its caustics and slows down its decay with the distance to
a slower rate $\sim1/R^{\xi}$ ($\xi<1/2$) compared with the $1/R^{1/2}$ decay
in uniform graphene, as a consequence of imperfect focusing away from $n=-1$,
e.g., for $\mathbf{R}_{2}$ on the cusp $(|nX_{1}|,0),$ we have $\xi
\approx0.24$ nearly independent of $E_{F}$, as shown in Fig.
\ref{G_PROPAGATOR}(c). By contrast, for $\mathbf{R}_{2}$ far away from the
caustics, the propagator recovers the $1/R^{1/2}$ decay of uniform graphene.
In addition to locally enhancing the propagator, the PNJ also slightly
decreases the propagation amplitude via the finite transmission $w(q_{y})$.
However, this effect is of minor importance because Klein tunneling
\cite{KatsnelsonNatPhys2006} allows carriers with a small incident angle
$\theta_{N}$ to go through the PNJ almost completely, as demonstrated by the
weak reflection in Fig. \ref{G_PROPAGATOR}(a) when the incident angle is small.

From the above analysis, it is clear that the PNJ qualitatively changes the
diverging spherical propagation of carriers in graphene into a converging one.
This greatly enhances the propagator near the caustics, so that its decay with
inter-spin distance $R$ slows down (for $E_{F}\neq0$) and even ceases (for
$E_{F}=0$). At large distances, the non-decaying propagator could lead to
$1/R$ decay of the RKKY interaction between two mirror symmetric spins about
the PNJ, as we demonstrate shortly (see Sec. IIIA).

Before concluding this subsection, we emphasize that the above analysis is
based on the $\mathbf{K}$-valley continuum model, featured by a single Dirac
cone and a circular Fermi contour. This simplified model ignores two important
effects:\ the trigonal warping at high energies and the presence of another
Dirac cone at $\mathbf{K}^{\prime}=-\mathbf{K}$. At high Fermi energies
$|E_{F}|\sim t$, the trigonal warping leads to non-circular Fermi contours,
while the inter-valley scattering may decrease the transmission probability
across a sharp PNJ \cite{LogemannPRB2015}. The former makes the focusing
effect no longer perfect even when $E_{F}=0$, while the latter decreases the
carrier propagator and hence the RKKY interaction. Even in the linear regime
$|E_{F}|\ll t$, the presence of two inequivalent valleys still gives rise to
inter-valley interference that significantly affect the RKKY interaction, as
discussed by Sherafati and Satpathy \cite{SherafatiPRB2011,SherafatiPRB2011a}
for uniform graphene. Here we briefly discuss this issue for the graphene PNJ.
First, we assume that the P-N interface does not induce inter-valley
scattering. Then, when both $\mathbf{K}$ and $\mathbf{K}^{\prime}$ valleys are
included, the 2$\times$2 carrier propagator from $\mathbf{R}_{1}$ to
$\mathbf{R}_{2}$ would be
\[
\mathbb{G}(\mathbf{R}_{2},\mathbf{R}_{1},E)=e^{i\mathbf{K}\cdot(\mathbf{R}%
_{2}-\mathbf{R}_{1})}\mathbf{g}(\mathbf{R}_{2},\mathbf{R}_{1}%
,E)+e^{i\mathbf{K}^{\prime}\cdot(\mathbf{R}_{2}-\mathbf{R}_{1})}%
\mathbf{g}^{\prime}(\mathbf{R}_{2},\mathbf{R}_{1},E),
\]
where $\mathbf{g}$ ($\mathbf{g}^{\prime}$) is the propagator in the presence
of the $\mathbf{K}$ ($\mathbf{K}^{\prime}$) valley alone [see Eq. (\ref{GF})
for the expression of $\mathbf{g}$]. The $\mathbf{K}^{\prime}$-valley
continuum model $\hat{h}^{\prime}=-v_{F}\boldsymbol{\hat{\sigma}}^{\ast}%
\cdot\mathbf{\hat{p}}+\mathrm{sgn}(x)V_{0}$ is the time reversal of the
$\mathbf{K}$-valley model [Eq. (\ref{HAMIL_CONTINUUM})], so that
$g_{s_{2}s_{1}}^{\prime}(\mathbf{R}_{2},\mathbf{R}_{1},E)=g_{s_{1}s_{2}%
}(\mathbf{R}_{1},\mathbf{R}_{2},E)$, i.e., the propagator of a $\mathbf{K}%
^{\prime}$-valley electron from the sublattice $s_{1}$ at $\mathbf{R}_{1}$ to
the sublattice $s_{2}$ at $\mathbf{R}_{2}$ is equal to the propagator of a
$\mathbf{K}$-valley electron from the sublattice $s_{2}$ at $\mathbf{R}_{2}$
back to the sublattice $s_{1}$ at $\mathbf{R}_{1}$. This also ensures the
time-reversal invariance of the total propagator:$\ \mathbb{G}_{s_{2}s_{1}%
}(\mathbf{R}_{2},\mathbf{R}_{1},E)=\mathbb{G}_{s_{1}s_{2}}(\mathbf{R}%
_{1},\mathbf{R}_{2},E)$. The RKKY\ interaction is given by Eq. (\ref{JS1S2})
with $G(\mathbf{R}_{2},\mathbf{R}_{1},E)$ replaced by the $(s_{\mathbf{R}_{2}%
},s_{\mathbf{R}_{1}})$ matrix element of $\mathbb{G}(\mathbf{R}_{2}%
,\mathbf{R}_{1},E)$. Consequently, the RKKY interaction consists of the
intra-valley contributions and the inter-valley interference term. The former
oscillates slowly on the length scale of the Fermi wave length, while the
latter oscillates rapidly as $e^{i(\mathbf{K-K}^{\prime})\cdot(\mathbf{R}%
_{2}-\mathbf{R}_{1})}$ on the atomic scale, similar to the case of uniform
graphene \cite{SherafatiPRB2011,SherafatiPRB2011a}. In the presence of
inter-valley scattering by the P-N interface, a quantitative description is
very difficult within the continuum model, but we still expect the
contribution from the inter-valley interference to be rapidly oscillating on
the atomic scale. Therefore, the slowly-varying \textit{envelope} of the
RKKY\ interaction is always determined by the \textit{intra-valley}
contributions, which are \textit{independent} of the direction of the P-N
interface with respect to the crystalline axis of graphene. In other words,
the continuum model suggests that the focusing of the RKKY\ interaction should
occur for an arbitrary direction of the P-N\ interface, as confirmed by our
subsequent numerical calculation based on the tight-binding model.

\subsection{Symmetry-protected non-oscillatory RKKY\ interaction}

For uniform graphene, the tight-binding Hamiltonian $\hat{H}_{0}$ [see Eq.
(\ref{H0})] possesses electron-hole symmetry $\hat{P}\hat{H}_{0}\hat{P}%
^{-1}=-\hat{H}_{0}$. \cite{SaremiPRB2007,BunderPRB2009,KoganPRB2011}, where
$\hat{P}$ inverts all the $\pi_{z}$-orbitals on sublattice $B$ but keeps all
$\pi_{z}$-orbitals on sublattice $A$ invariant, i.e., $\hat{P}|\mathbf{R}%
\rangle=\pm|\mathbf{R}\rangle$, with the upper (lower) sign for $s_{\mathbf{R}%
}=A$ ($s_{\mathbf{R}}=B$). For \textit{undoped} graphene, this makes the
RKKY\ interaction between local spins on the same (opposite) sublattice always
ferromagnetic (antiferromagnetic), irrespective of their distance. However,
once the graphene is doped, the RKKY interaction recovers its
\textquotedblleft universal\textquotedblright\ oscillation with a
characteristic wavelength $\lambda_{F}/2$ ($\lambda_{F}$ is the Fermi
wavelength) between ferromagnetic and anti-ferromagnetic couplings, as also
found in many other materials.

For the graphene PNJ, the presence of the junction potential breaks the
electron-hole symmetry of uniform graphene. Moreover, since both the N region
and the P region are doped, the RKKY interaction is also expected to oscillate
between ferromagnetic and antiferromagnetic couplings with the
distance.\textbf{ }Interestingly, we find that the electron-hole symmetry can
be restored under certain conditions. Let us consider a general graphene PNJ
described by the tight-binding Hamiltonian Eq. (\ref{HAMIL_TB}) with a general
on-site junction potential $V_{\mathbf{R}}$ and define the mirror reflection
operator $\hat{M}$ that maps the $\pi_{z}$-orbital $|\mathbf{R}\rangle$ to
another $\pi_{z}$-orbital $|\mathbf{R}_{\mathrm{m}}\rangle$ at the mirror
image location $\mathbf{R}_{\mathrm{m}}$ about the P-N interface, i.e.,
$\hat{M}|\mathbf{R}\rangle=|\mathbf{R}^{\mathrm{m}}\rangle$. The key
observation is that as long as the mirror reflection $\hat{M}$ about the P-N
interface keeps the graphene lattice invariant but inverts the junction
potential (i.e., $V_{\mathbf{R}}=-V_{\mathbf{R}^{\mathrm{m}}}$), the PNJ
Hamiltonian $\hat{H}$ possesses a generalized electron-hole symmetry:
$(\hat{P}\hat{M})\hat{H}(\hat{P}\hat{M})^{-1}=-\hat{H}$. This ensures the
eigen-energies of the PNJ to appear in pairs $(\varepsilon,-\varepsilon)$ and
the corresponding eigenstates $|\phi_{\varepsilon}\rangle$ and $|\phi
_{-\varepsilon}\rangle$ obey $|\phi_{-\varepsilon}\rangle=\hat{P}\hat{M}%
|\phi_{\varepsilon}\rangle$, similar to the electron-hole symmetry in uniform
graphene \cite{SaremiPRB2007,BunderPRB2009,KoganPRB2011}. As a consequence of
this symmetry, when $E_{F}=0$, the Matsubara Green's function $\mathcal{G}%
(\mathbf{R},\mathbf{R}^{\prime},\tau)$ \cite{MahanBook2000} obeys
$\mathcal{G}(\mathbf{R}_{1},\mathbf{R}_{1}^{\mathrm{m}},-\tau)=\pm
\mathcal{G}(\mathbf{R}_{1}^{\mathrm{m}},\mathbf{R}_{1},\tau)$, with the upper
(lower) sign for $\mathbf{R}_{1}$ and $\mathbf{R}_{1}^{\mathrm{m}}$ on the
opposite (same) sublattices. The time-reversal symmetry further dictates the
Matsubara Green's functions to be real. According to the imaginary-time
formalism for the RKKY\ interaction
\cite{SaremiPRB2007,BunderPRB2009,KoganPRB2011} [equivalent to the real-time
formalism in Eq. (\ref{JS1S2})], the sign of the RKKY\ range function is
determined by $\mathcal{G}(\mathbf{R}_{1},\mathbf{R}_{2},-\tau)\mathcal{G}%
(\mathbf{R}_{2},\mathbf{R}_{1},\tau)$. Therefore, when the two local spins are
mirror symmetric about the PNJ, i.e., $\mathbf{R}_{2}=\mathbf{R}%
_{1}^{\mathrm{m}}$, their RKKY interaction is always ferromagnetic
(antiferromagnetic) on the same (opposite) sublattices, irrespective of their
distance. If the P-N interface is along the zigzag direction, then
$\mathbf{R}_{1}$ and $\mathbf{R}_{1}^{\mathrm{m}}$ are always on opposite
sublattices, so the RKKY\ interaction is antiferromagnetic. If the P-N
interface is along the armchair direction, then $\mathbf{R}_{1}$ and
$\mathbf{R}_{1}^{\mathrm{m}}$ are always on the same sublattice, so the
RKKY\ interaction is ferromagnetic.

\section{Numerical results}

\begin{figure}[ptbh]
\centering
\subfigure{}{
\begin{minipage}{\columnwidth}
\centering
\includegraphics[width=\columnwidth,clip]{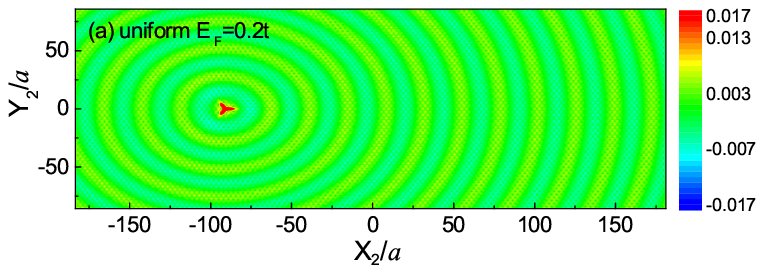}  \\
\end{minipage}
} \subfigure{}{
\begin{minipage}{\columnwidth}
\centering
\includegraphics[width=\columnwidth,clip]{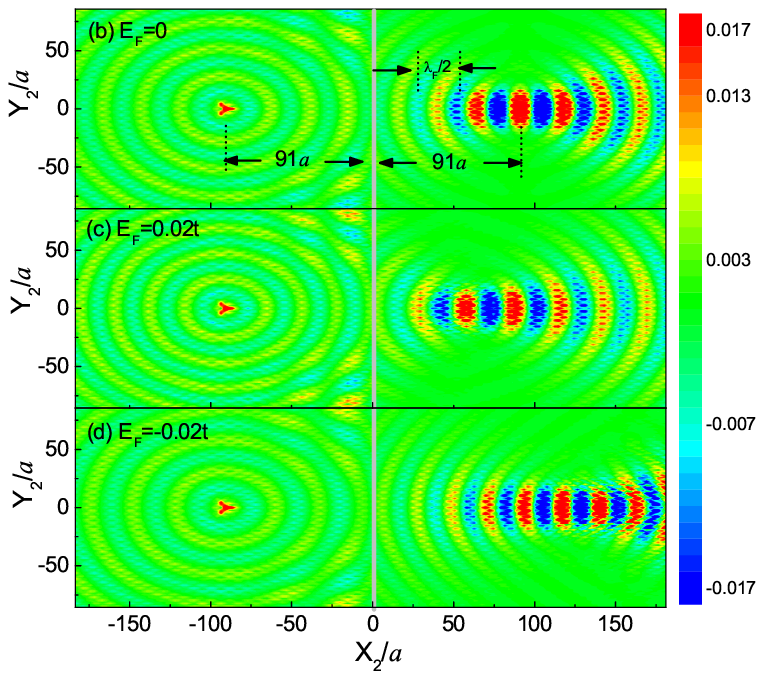}  \\
\end{minipage}
} \caption{Scaled RKKY range function $\mathcal{J}R^{2}/a^{2}$ vs.\textbf{
}$\mathbf{R}_{2}=(X_{2},Y_{2})$\textbf{ }on the $B$ sublattice for
fixed\textbf{ }$\mathbf{R}_{1}=(-91a,0)$\textbf{ }on the $A$ sublattice in (a)
uniform graphene with electron doping $E_{F}=0.2t$ and (b)-(d) graphene PNJ
with junction potential $V_{0}=0.2t$\textbf{ }and different Fermi energies.
The same color scale is used for all the panels\textbf{, }i.e., blue (red) for
negative (positive) or equivalently ferromagnetic (antiferromagnetic)
RKKY\ interactions. }%
\label{G_CONTOUR}%
\end{figure}

Here we calculate the RKKY interaction in the graphene PNJ numerically based
on the tight-binding model [Eq. (\ref{HAMIL_TB})], with the on-site junction
potential $V_{\mathbf{R}}=-V_{0}$ ($V_{\mathbf{R}}=+V_{0}$) in the N (P)
region. For convenience, we introduce the dimensionless RKKY range function
$\mathcal{J}\equiv{tJ/J_{0}^{2}}$. Due to the transformation $\hat
{H}\rightarrow-\hat{H}$ upon $(V_{0},t)\rightarrow(-V_{0},-t)$ and the
time-reversal symmetry, $\mathcal{J}$ is invariant upon $(E_{F},V_{0}%
)\rightarrow(-E_{F},-V_{0})$ (see Appendix B), so we need only consider
$V_{0}>0$. Our results show that for different orientations of the PNJ (e.g.,
along the zigzag direction, the armchair direction, and a slightly misaligned
direction) and different sublattice locations of $\mathbf{R}_{1}$ and
$\mathbf{R}_{2}$, the RKKY interaction exhibits similar anomalous focusing
behaviors, consistent with our previous analysis based on the continuum model
in Sec. IIA. For specificity, we present our results for a PNJ along the
zigzag direction and \textit{always} take $\mathbf{R}_{1}$ on the $A$
sublattice and $\mathbf{R}_{2}$ on the $B$ sublattice.

\subsection{Anomalous focusing of RKKY\ interaction}

For the first local spin $\hat{\mathbf{S}}_{1}$ fixed at $\mathbf{R}%
_{1}=(-91a,0)$ [$a$ is the C-C bond length, see Fig. \ref{G_PNJ}(a)], the
spatial map of the scaled range function $\mathcal{J}R^{2}/a^{2}$ as a
function of the location $\mathbf{R}_{2}=(X_{2},Y_{2})$ of the second local
spin $\hat{\mathbf{S}}_{2}$ is shown in Fig. \ref{G_CONTOUR}(a) for uniform
graphene and Fig. \ref{G_CONTOUR}(b)-(d) for graphene PNJ. Here we follow Ref.
\onlinecite{LeePRB2012} and use the multiplication factor $R^{2}/a^{2}$ to
remove the intrinsic decay $\propto1/R^{2}$ of the RKKY interaction in uniform
graphene \cite{BreyPRL2007,SherafatiPRB2011a}. This helps us to present an
overall view of the spatial texture of the RKKY interaction over both the N
region and the P region in a single contour plot and highlights the focusing
of the RKKY\ interaction by the P-N interface. For example, in uniform
graphene with $V_{0}=0$ and $E_{F}=0.2t$ [Fig. \ref{G_CONTOUR}(a)], the scaled
range function does not decay, manifesting the intrinsic $1/R^{2}$ decay of
the RKKY interaction. By contrast, in the graphene PNJ [Fig. \ref{G_CONTOUR}%
(b)-(d)], the P-N interface induces strong local enhancement of the RKKY
interaction in the P region, but it has a negligible influence in the N
region. This is an obvious consequence of anomalous focusing: in the N region,
the carrier propagation remains diverging, similar to uniform graphene shown
by Fig. \ref{G_CONTOUR}(a); while in the P region, the carrier wave is
refocused by the PNJ \cite{CheianovScience2007}. For $E_{F}=0$ in Fig.
\ref{G_CONTOUR}(b), corresponding to an effective refraction index $n=-1$, the
RKKY interaction is significantly enhanced when $\hat{\mathbf{S}}_{2}$ locates
near the mirror image of $\hat{\mathbf{S}}_{1}$ about the PNJ. For
$E_{F}=0.02t$ in Fig. \ref{G_CONTOUR}(c) [$E_{F}=-0.02t$ in Fig.
\ref{G_CONTOUR}(d)], corresponding to $n\approx\allowbreak-0.82$
($n\approx-1.2$), the maximum of the RKKY interaction shifts towards (away
from) the PNJ, consistent with the shift of the caustics [see Eq.
(\ref{CAUSTICS})].

Now we discuss two fine features in Fig. \ref{G_CONTOUR}. First, for uniform
graphene, Fig. \ref{G_CONTOUR}(a) reproduces the C$_{3\mathrm{v}}$ spatial
symmetry at short distances \cite{LeePRB2012}, the slow oscillations with a
characteristic wavelength $\lambda_{F}/2$
\cite{BreyPRL2007,SherafatiPRB2011a,LeePRB2012}, and the rapid oscillations on
the atomic scale\textbf{ }due to the inter-valley interference
\cite{SaremiPRB2007,SherafatiPRB2011a,LeePRB2012}, as described by the
dimensionless range function at large distances $(q_{F}R\gg1)$
\cite{SherafatiPRB2011a}:%
\begin{equation}
\mathcal{J}_{\mathrm{uniform}}\approx\frac{q_{F}a}{(R/a)^{2}}\frac{9}%
{32\pi^{2}}\{1-\cos\left[  (\mathbf{K}-\mathbf{K}^{\prime})\cdot
\mathbf{R}-2\theta_{\mathbf{R}}\right]  \}\sin(2q_{F}R),\label{AGF}%
\end{equation}
where $q_{F}$ is the Fermi momentum and $\theta_{\mathbf{R}}$ is the angle
between $\mathbf{R}\equiv\mathbf{R}_{2}-\mathbf{R}_{1}$ and $\mathbf{K-K}%
^{\prime}$. For the graphene PNJ in Fig. \ref{G_CONTOUR}(b)-(d), the
RKKY\ interaction also consists of a slowly-varying envelope and a
rapidly-varying part that oscillates on the atomic scale. As discussed at the
end of Sec. IIA, the former comes from the intra-valley contributions, while
the latter comes from the inter-valley interference and hence oscillates with
a momentum $\mathbf{K-K}^{\prime}$, similar to Eq. (\ref{AGF}). Notice that in
Figs. \ref{G_CONTOUR}(a)-(d), the most rapid atomic-scale oscillation occurs
along the $y$ axis [i.e., the zigzag direction, see Fig. \ref{G_PNJ}(a)],
consistent with previous studies in uniform graphene
\cite{SherafatiPRB2011,SherafatiPRB2011a}.

Second, in Fig. \ref{G_CONTOUR}(b)-(d), there is no local enhancement of the
RKKY\ interaction near the P-N interface. According to Eq. (\ref{JS1S2}), this
manifests the fact that there is no local charge accumulation near the P-N
interface, since the incident electron wave emanating from $\mathbf{R}_{1}$
either reflects back or transmits through the P-N interface. The interference
between the incident wave and the reflection wave in the N\ region (near the
P-N interface) can be clearly seen by comparing Fig. \ref{G_CONTOUR}(b)-(d) to
Fig. \ref{G_CONTOUR}(a).

\begin{figure}[t]
\includegraphics[width=\columnwidth]{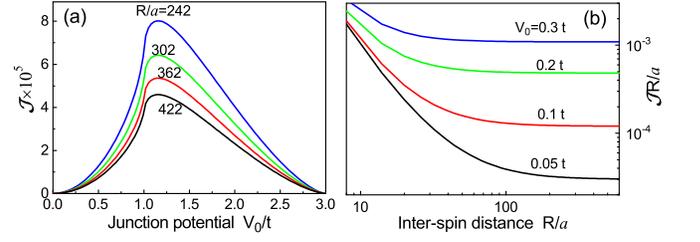}\caption{{RKKY range function
between two local spins mirror symmetric about the PNJ with $E_{F}=0$.} (a)
$\mathcal{J}$ vs. junction potential for different inter-spin distances $R$.
(b) $\mathcal{J}R/a$ vs. $R$ for different junction potentials.}%
\label{G_SCALING}%
\end{figure}

Let us consider the RKKY interaction between two mirror symmetric spins in
graphene PNJ at $E_{F}=0$, i.e., electron doping $V_{0}$ in the N region and
hole doping $V_{0}$ in the P region. According to the symmetry analysis in
Sec. IIB, for the PNJ with its interface along the zigzag direction, the RKKY
interaction between two mirror symmetric spins is always antiferromagnetic.
This feature is demonstrated by Fig. \ref{G_SCALING}. As shown in Fig.
\ref{G_SCALING}(a), in the linear regime ($V_{0}\ll t$), the RKKY interaction
$\mathcal{J}\propto V_{0}^{2}$ increases quadratically with $V_{0}$. As shown
in Fig. \ref{G_SCALING}(b), the scaled RKKY interaction strength
$\mathcal{J}R/a$ is $R$-independent at large distances. This indicates that
$\mathcal{J}$ follows $1/R$ asymptotic decay due to the perfect refocusing, in
sharp contrast to the $1/R^{d}$ asymptotic decay in a great diversity of doped
$d$-dimensional materials, as well as the $1/R^{3}$ asymptotic decay in
undoped graphene. Therefore, the RKKY\ interaction at $V_{0}\ll t$ and
$R\gg\lambda_{F}$ can be well approximated by the analytical expression
$\mathcal{J}\approx0.012{(V_{0}/t)^{2}}/({R/a}),$ where the constant $0.012$
is obtained by fitting the data in Fig. \ref{G_SCALING}(a)-(b). For
comparison, in uniform graphene with the same doping level as the PNJ, the
envelope of the RKKY interaction [see Eq. (\ref{AGF})] has the asymptotic form
$\mathcal{J}_{\mathrm{uniform}}\approx0.037({V_{0}/t})/({R^{2}/a^{2}})$. The
RKKY\ interaction in the graphene PNJ differs qualitatively from that in
uniform graphene in the scaling with both the distance ($\mathcal{J}%
_{\mathrm{uniform}}\propto1/R^{2}$ vs. $\mathcal{J}\propto1/R$) and the
junction potential $V_{0}$ or equivalently carrier concentration
($\mathcal{J}_{\mathrm{uniform}}\propto V_{0}$ vs. $\mathcal{J}\propto
V_{0}^{2}$). Since the localized spins are usually fixed, dynamic tuning of
the PNJ by electric gating potentially allows for selective control of
localized spins, an important ingredient for spin-based quantum computation.

\subsection{Experimental feasibility and generalization to other materials}

Since the focusing of the RKKY\ interaction is dominated by the contribution
from the electron states near the Fermi level, and these states cannot not
\textquotedblleft feel\textquotedblright\ any potential variation on the
length scale $\ll$ Fermi wavelength $\lambda_{F}$, the finite width of the
PNJ\ has a small influence as long as it is much smaller than $\lambda_{F}$.
Taking $V_{0}=0.1t$ and the first local spin at $\mathbf{R}_{1}=(-151a,0)$ as
an example, using the experimentally fabricated linear PNJ
\cite{WilliamsNatNano2011} of width $29a\approx4.1$ nm, instead of a sharp
PNJ, only reduces the magnitude of the RKKY\ interaction by $\sim25\%$ without
changing the $1/R$ asymptotic scaling.

In the presence of a finite gap $\Delta$\ (e.g., due to substrate mismatch
\cite{HuntScience2013}) in the Dirac spectrum of graphene, as long as
$\Delta\ll|V_{0}|$, the gap does not significantly influence the states near
the Fermi level, which dominates the anomalous focusing effect. This has been
confirmed by our numerical calculation using $E_{F}=0$, $V_{0}=0.2t$, and a
typical gap $\Delta=0.03t$: no appreciable change of the focusing behavior
occurs. As a matter of fact, from our analytical analysis following Eq.
(\ref{GF}), it is clear that perfect focusing of the PNJ with $E_{F}=0$
essentially arises from the opposite momenta $q_{N,x}=-q_{P,x}=[(V_{0}%
/v_{F})^{2}-q_{y}^{2}]^{1/2}$ in the N region and P region of the PNJ, which
suppresses the rapidly oscillating phase $\mathbf{q}_{P}\cdot\mathbf{R}%
_{2}-\mathbf{q}_{N}\cdot\mathbf{R}_{1}=0$ [see Eq. (\ref{GF})] as long as
$\mathbf{R}_{1}$ and $\mathbf{R}_{2}$ are mirror symmetric about the PNJ. The
key ingredients of this effect are the circular Fermi contours and the
opposite group velocities in the N region and P region, although the linear
dispersion and gapless feature of graphene allows high transmission of
electron waves (i.e., Klein tunneling) and hence quantitatively stronger
focusing effect. Consequently, similar principle should lead to similar effect
in other materials with a nonlinear dispersion and/or a finite gap (e.g.,
silicene PNJ \cite{YamakagePRB2013}). For example, we have numerically
verified that the local enhancement of the RKKY interaction remains effective
even when the gap of the graphene PNJ\ increases to $\Delta=0.1$ $t$.

The anomalous focusing is pronounced at low temperature and persists up to
nitrogen temperature \cite{CheianovScience2007}. To observe the $1/R$
long-range RKKY interaction across the graphene PNJ, experiments should be
carried out at a temperature higher than the Kondo temperature to avoid the
screening of the local spins by the carriers \cite{UchoaPRL2011}. The
anomalous focusing in graphene PNJ has been demonstrated by two recent
experiments \cite{LeeNatPhys2015,ChenScience2016}. Thus we expect our
theoretical results to be experimentally accessible.

\section{Summary}

As opposed to previous works that explore the influence of electronic states
and energy band structures of uniform\textbf{ }2D systems on the
carrier-mediated RKKY interaction, we have proposed a very different way to
manipulate the RKKY\ interaction: tailoring the carrier propagation and
interference via a well-known electron optics phenomenon in graphene P-N
junctions -- the anomalous focusing effect. This gives rise to rich spatial
interference patterns and locally enhanced, symmetry-protected non-oscillatory
RKKY\ interaction, which may pave the way towards long-range spin-spin
interaction for scalable graphene-based spintronics devices. The key physics
leading to the focusing of the RKKY\ interaction is the focusing of the
carrier spin fluctuation emanating from a local spin. In this context, we
notice a very relevant work by Guimaraes \textit{et al. }%
\cite{GuimaraesJPC2011}, which shows that a gate-defined curved boundary in
graphene can focus the spin current emanating from a precessing magnetic
moment onto a specific point. We expect that this spin current lens could be
utilized as an alternative way to focus the RKKY\ interaction.

\section*{Acknowledgements}

This work was supported by the MOST of China (Grant No. 2014CB848700 and No.
2015CB921503), the NSFC (Grant No. 11274036, No. 11322542, No. 11434010, No.
11404043 and No. 11504018), and the NSFC program for \textquotedblleft
Scientific Research Center\textquotedblright\ (Grant No. U1530401). We
acknowledge the computational support from the Beijing Computational Science
Research Center (CSRC). J. J. Z. thanks the new research direction support
program of CQUPT.

\appendix

\section{Propagator in $\mathbf{K}$-valley continuum model}

Here we derive the 2$\times$2 matrix propagator $\mathbf{g}(\mathbf{r}%
,\mathbf{r}^{\prime},E)$ with $E\in\lbrack-V_{0},V_{0}]$ in the $\mathbf{K}%
$-valley continuum model [Eq. (\ref{HAMIL_CONTINUUM}) of the main text]. Due
to translational invariance along the $y$ axis, the 2D propagator
\begin{equation}
\mathbf{g}(\mathbf{r},\mathbf{r}^{\prime},E)\equiv\int\frac{dq_{y}}{2\pi
}e^{iq_{y}(y-y^{\prime})}\mathbf{g}_{\mathrm{1D}}(x,x^{\prime},E)
\label{G2D_A}%
\end{equation}
is determined by the 1D propagator $\mathbf{g}_{\mathrm{1D}}(x,x^{\prime},E)$
of the 2$\times$2 Hamiltonian $\mathbf{h}_{\mathrm{1D}}(x,-i\partial
_{x})=-iv_{F}\hat{\sigma}_{x}\partial_{x}+v_{F}\hat{\sigma}_{y}q_{y}%
+\mathrm{sgn}(x)V_{0}$, with the dependence of $\mathbf{g}_{\mathrm{1D}%
}(\cdots)$ and $\mathbf{h}_{\mathrm{1D}}(\cdots)$ on $q_{y}$ omitted for
brevity. Here $\mathbf{g}_{\mathrm{1D}}(x,x^{\prime},E)$ obeys the
differential equations
\begin{align*}
\lbrack E+i0^{+}-\mathbf{h}_{\mathrm{1D}}(x,-i\partial_{x})]\mathbf{g}%
_{\mathrm{1D}}(x,x^{\prime},E)  &  =\delta(x-x^{\prime}),\\
\mathbf{g}_{\mathrm{1D}}(x,x^{\prime},E)[E+i0^{+}-\mathbf{h}_{\mathrm{1D}%
}(x^{\prime},-i\overleftarrow{\partial_{x^{\prime}}})]  &  =\delta
(x-x^{\prime}),
\end{align*}
($\overleftarrow{\partial_{x^{\prime}}}$ acting on the left) and the
continuity conditions
\begin{align*}
\mathbf{g}_{\mathrm{1D}}(x+0^{+},x,E)-\mathbf{g}_{\mathrm{1D}}(x-0^{+},x,E)
&  =-i\sigma_{x}/v_{F},\\
\mathbf{g}_{\mathrm{1D}}(x,x+0^{+},E)-\mathbf{g}_{\mathrm{1D}}(x,x-0^{+},E)
&  =i\sigma_{x}/v_{F}.
\end{align*}
Then $\mathbf{g}_{\mathrm{1D}}(x,x^{\prime},E)$ is obtain by first calculating
the general solutions in the region $x\neq x^{\prime}$ and then matching them
using the boundary conditions.

To present the results in a physically intuitive way, we introduce the
following concepts. Given the energy $E$ and momentum $q_{y}$, there is one
right-going eigenstate $|+,\mathbf{q}_{N}\rangle$ with $\mathbf{q}_{N}%
\equiv(q_{N,x},q_{y})$ and one left-going eigenstate $|+,\mathbf{\tilde{q}%
}_{N}\rangle$ with $\mathbf{\tilde{q}}_{N}\equiv(-q_{N,x},q_{y})$ in the N
region, as well as one right-going eigenstate $|-,\mathbf{q}_{P}\rangle$ with
$\mathbf{q}_{P}\equiv(q_{P,x},q_{y})$ and one left-going eigenstate
$|-,\mathbf{\tilde{q}}_{P}\rangle$ with $\mathbf{\tilde{q}}_{P}\equiv
(-{q}_{P,x},q_{y})$ in the P region, where $|s,\mathbf{q}\rangle\propto
e^{i\mathbf{q}\cdot\mathbf{r}}|u_{s}(\mathbf{q})\rangle$ is the eigenstate of
uniform graphene in the conduction band $(s=+)$ or valence band $(s=-$). For
small $|q_{y}|$, we choose $q_{N,x}>0$ and $q_{P,x}<0$, so that $|+,\mathbf{q}%
_{N}\rangle$ and $|-,\mathbf{q}_{P}\rangle$ ($|+,\mathbf{\tilde{q}}_{N}%
\rangle$ and $|-,\mathbf{\tilde{q}}_{P}\rangle$) propagate from the left
(right)\ to the right (left) without decay. For large $|q_{y}|$, we choose
$\operatorname{Im}q_{N,x}>0$ and $\operatorname{Im}q_{P,x}>0$, so that
$|+,\mathbf{q}_{N}\rangle$ and $|-,\mathbf{q}_{P}\rangle$ ($|+,\mathbf{\tilde
{q}}_{N}\rangle$ and $|-,\mathbf{\tilde{q}}_{P}\rangle$) decays to zero at
$x\rightarrow+\infty$ $(x\rightarrow-\infty)$.

In terms of these left-going and right-going eigenstates with given energy $E$
and $q_{y}$, the 1D propagator from $x^{\prime}$ in the N region to $x$ in the
P\ region is%
\begin{equation}
\mathbf{g}_{\mathrm{1D}}(x,x^{\prime},E)=\frac{w(q_{y})}{iv_{x}(\mathbf{q}%
_{N})}|u_{-}(\mathbf{q}_{P})\rangle\langle u_{+}(\mathbf{q}_{N})|e^{i(q_{P,x}%
x-q_{N,x}x^{\prime})}, \label{G1D_A}%
\end{equation}
where $w(q_{y})=2\cos\theta_{N}/(e^{-i\theta_{N}}+e^{i\theta_{P}})$ is the
transmission coefficient and $\mathbf{v}(\mathbf{q})\equiv v_{F}%
\mathbf{q}/|\mathbf{q}|$ is the group velocity, with $\theta_{N}$ ($\theta
_{P}$) is the incident (transmission) angle defined via $v_{F}(q_{N,x}%
+iq_{y})=(E+V_{0})e^{i\theta_{N}}$ and $v_{_{F}}(q_{P,x}+iq_{y})=-(V_{0}%
-E)e^{i\theta_{P}}$. Substituting Eq. (\ref{G1D_A}) into Eq. (\ref{G2D_A})
gives the 2D propagator $\mathbf{g}(\mathbf{r},\mathbf{r}^{\prime
},E)|_{\mathbf{r}^{\prime}\in\mathrm{N},\mathbf{r}\in\mathrm{P}}$ in Eq.
(\ref{GF}) of the main text. For $x^{\prime}<x<0$, the 2D propagator
$\mathbf{g}(\mathbf{r},\mathbf{r}^{\prime},E)$ is the sum of the direct,
forward propagation from $\mathbf{r}^{\prime}$ to $\mathbf{r}$,
\[
\int\frac{dq_{y}}{2\pi}\frac{|u_{+}(\mathbf{q}_{N})\rangle\langle
u_{+}(\mathbf{q}_{N})|}{iv_{x}(\mathbf{q}_{N})}e^{i\mathbf{q}_{N}%
\cdot(\mathbf{r}-\mathbf{r}^{\prime})},
\]
and the contribution from the reflected wave
\[
\int\frac{dq_{y}}{2\pi}r(q_{y})\frac{|u_{+}(\mathbf{\tilde{q}}_{N}%
)\rangle\langle u_{+}(\mathbf{q}_{N})|}{iv_{x}(\mathbf{q}_{N})}%
e^{i\mathbf{\tilde{q}}_{N}\cdot\mathbf{r}}e^{-i\mathbf{q}_{N}\cdot
\mathbf{r}^{\prime}}%
\]
via three steps: the propagation from $\mathbf{r}^{\prime}$ to the PNJ, the
reflection by the PNJ, and the propagation of the reflected wave from the PNJ
to $\mathbf{r}$. For $x<x^{\prime}<0$, $\mathbf{g}(\mathbf{r},\mathbf{r}%
^{\prime},E)$ is obtained from $\mathbf{g}(\mathbf{r},\mathbf{r}^{\prime
},E)|_{x^{\prime}<x<0}$ by replacing the direct, forward propagation by the
direct, backward propagation:%
\[
\int\frac{dq_{y}}{2\pi}\frac{|u_{+}(\mathbf{\tilde{q}}_{N})\rangle\langle
u_{+}(\mathbf{\tilde{q}}_{N})|}{iv_{x}(\mathbf{q}_{N})}e^{i\mathbf{\tilde{q}%
}_{N}\cdot(\mathbf{r}-\mathbf{r}^{\prime})}.
\]

\section{Symmetry of propagators and RKKY interaction in tight-binding model}

In the tight-binding model, the propagator from $\mathbf{R}^{\prime}$ to $\mathbf{R}$ is $G(\mathbf{R},\mathbf{R}^{\prime},E)\equiv\langle\mathbf{R}%
|(E-\hat{H}+i0^{+})^{-1}|\mathbf{R}^{\prime}\rangle$. Using the invariance
$\hat{\theta}\hat{H}\hat{\theta}^{-1}=\hat{H}$ and $\hat{\theta}%
|\mathbf{R}\rangle=|\mathbf{R}\rangle$ under time-reversal transformation, we
have $G(\mathbf{R},\mathbf{R}^{\prime},E)\equiv\langle\mathbf{R}|(E-\hat
{H}-i0^{+})^{-1}|\mathbf{R}^{\prime}\rangle^{\ast}=G(\mathbf{R}^{\prime
},\mathbf{R},E)$. Inverting the nearest-neighbor hopping amplitude
$t\rightarrow-t$ amounts to the transformation $|\mathbf{R}\rangle
\rightarrow\mathrm{sgn}(\mathbf{R})|\mathbf{R}\rangle$ in the Hamiltonian and
hence $G(\mathbf{R},\mathbf{R}^{\prime},E)|_{t\rightarrow-t}=\mathrm{sgn}%
(\mathbf{R})\mathrm{sgn}(\mathbf{R}^{\prime})G(\mathbf{R},\mathbf{R}^{\prime
},E)$, where $\mathrm{sgn}(\mathbf{R})=+1$ if $\mathbf{R}$ locates on the $A$
sublattice and $\mathrm{sgn}(\mathbf{R})=-1$ if $\mathbf{R}$ locates on the
$B$ sublattice. Since $\hat{H}\rightarrow-\hat{H}$ under $(V_{0}%
,t)\rightarrow(-V_{0},-t)$ and $\hat{\theta}\hat{H}\hat{\theta}^{-1}=\hat{H}$,
we have
\begin{align*}
G(\mathbf{R},\mathbf{R}^{\prime},E)  &  \rightarrow-\langle\mathbf{R}%
|(E-\hat{H}-i0^{+})^{-1}|\mathbf{R}^{\prime}\rangle\\
&  =-\langle\mathbf{R}|(E-\hat{H}+i0^{+})^{-1}|\mathbf{R}^{\prime}%
\rangle^{\ast}\\
&  =-G^{\ast}(\mathbf{R},\mathbf{R}^{\prime},E)
\end{align*}
under $(E,V_{0},t)\rightarrow(-E,-V_{0},-t)$. Combining the transformation
properties of the propagator under $t\rightarrow-t$ and $(E,V_{0}%
,t)\rightarrow(-E,-V_{0},-t)$ gives $G(\mathbf{R},\mathbf{R}^{\prime
},E)\rightarrow-\mathrm{sgn}(\mathbf{R})\mathrm{sgn}(\mathbf{R}^{\prime
})G^{\ast}(\mathbf{R},\mathbf{R}^{\prime},E)$ upon $(E,V_{0})\rightarrow
(-E,-V_{0})$. Using the generalized electron-hole symmetry of the graphene
PNJ, we also have $G(\mathbf{R},\mathbf{R}^{\mathrm{m}},E)=-\mathrm{sgn}%
(\mathbf{R})\mathrm{sgn}(\mathbf{R}^{\mathrm{m}})G^{\ast}(\mathbf{R}%
,\mathbf{R}^{\mathrm{m}},-E)$.

The above transformation properties of the propagator leads to the
corresponding properties of the RKKY range function $J$, e.g., $J$ is
invariant upon $t\rightarrow-t$. Similarly, upon $(V_{0},E_{F})\rightarrow
(-V_{0},-E_{F})$, we have
\begin{align*}
J  &  \rightarrow-\frac{J_{0}^{2}}{2\pi}\int_{E_{F}}^{+\infty}%
dE\operatorname{Im}G^{2}({\mathbf{R}_{2}},{\mathbf{R}_{1}},-E)|_{V_{0}%
\rightarrow-V_{0}}\\
&  =\frac{J_{0}^{2}}{2\pi}\int_{E_{F}}^{+\infty}dE\operatorname{Im}%
G^{2}({\mathbf{R}_{2}},{\mathbf{R}_{1}},E).
\end{align*}
Using $\int_{-\infty}^{\infty}\operatorname{Im}G^{2}({\mathbf{R}_{2}%
},{\mathbf{R}_{1}},E)dE=0$, we see that $J$ is invariant upon $(V_{0}%
,E_{F})\rightarrow(-V_{0},-E_{F})$. Also, when the two spins locate at mirror
symmetric points $\mathbf{R}$ and $\mathbf{R}^{\mathrm{m}}$, their RKKY
interaction $J$ is invariant upon $E_{F}\rightarrow-E_{F}$.


\end{document}